\documentclass[3p,times]{elsarticle}
\usepackage{ecrc}
\volume{00}
\firstpage{1}
\journalname{Nuclear Physics A}
\runauth{R.Shyam et al.}
\jid{NUPHA}
\jnltitlelogo{Nuclear Physics A}
\CopyrightLine{2012}{Published by Elsevier Ltd.}
\usepackage{amssymb}
\usepackage[figuresright]{rotating}

\begin{document}

\begin{frontmatter}

\dochead{}

\title{$\Xi^-$ hyperon and hypernuclear production in the $(K^-,K^+)$ 
reaction on nucleon and nuclei in a field theoretical model}
\author[a]{R. Shyam}

\address[a]{Saha Institute of Nuclear Physics, 1/AF Bidhan Nagar, Kolkata 700064, 
India}

\begin{abstract}
We investigate the production of a cascade hyperon ($\Xi$) and bound $\Xi$ 
hypernuclei in the $(K^-,K^+)$ reaction on proton and nuclear targets, respectively, 
within a covariant effective Lagrangian model. The $K^+\Xi^-$ production vertex is 
described by excitation, propagation and decay of $\Lambda$ and $\Sigma$ resonance 
states in the initial collision of a $K^-$ meson with a free or bound  proton in the 
incident channel. The parameters of the resonance vertices are taken from previous 
studies and SU(3) symmetry considerations. The model is able to provide a good 
description of the available data on total and differential cross sections for the 
$p(K^-, K^+)\Xi^-$ reaction. The same mechanism was used to describe the hypernuclear
production reactions $^{12}$C$(K^-,K^+)^{12}{\!\!\!_{\Xi^-}}$Be and 
$^{28}$Si$(K^-,K^+)^{28}{\!\!\!_{\Xi^-}}$Mg, where $\Xi$ bound state spinors 
calculated within a phenomenological model have been employed. Both the elementary 
and hypernuclear production cross sections are dominated by the contributions from 
the $\Lambda$(1520) intermediate resonant state. The beam momentum dependence of 
the $0^\circ$ differential cross sections for the formation of the $\Xi$ hypernuclei 
is found to be remarkably different from what has been observed previously in the 
impulse approximation model calculations.
\end{abstract}


\end{frontmatter}
\section{Introduction}
\label{1}

Spectroscopy of hadrons is one of the key tools for studying quantum chromodynamics 
(QCD) in the non-perturbative regime. Lattice simulations, which provide the only 
{\it ab initio} calculations of QCD in this regime, are now able to reproduce a large
part of the experimentally observed ground state hadron spectrum. This development is 
important particularly for those processes which are difficult to explore in the 
laboratory, such as hyperon-hyperon and hyperon-nucleon interactions that may 
affect the late stages of the supernova evolution. The study of the double 
strangeness ($S$) hypernuclei is of great importance in this context. The binding 
energies and widths of the $\Xi$ hypernuclear states are likely to determine the 
strengths of the $\Xi N$ and $\Xi N \to \Lambda \Lambda$ interactions. This basic 
information is key to testing the quark exchange aspect of the strong interaction 
because long range pion exchange plays essentially a very minor role in the $S = -2$ 
sector. This input is also vital for understanding the multi-strange hadronic or 
quark matter. Since strange quarks are negatively charged they are preferred in charge 
neutral dense matter. Thus these studies are of crucial value for investigating the 
role of strangeness in the equation of state at high density, as probed in the 
cores of neutron stars~\cite{bie10}. 
 
The $(K^-,K^+)$ reaction leads to the transfer of two units of both charge and 
strangeness to the target nucleus. Thus this reaction is one of the most promising 
ways of studying the $S = -2$ systems such as $\Xi$ hypernuclei and a dibaryonic 
resonance ($H$), which is a near stable six-quark state with spin parity of $0^+$ 
and isospin 0~\cite{jaf77,mul83,bea11}. The $(K^-,K^+)$ reaction implants a $\Xi$ 
hyperon in the nucleus through the elementary $p(K^-, K^+)\Xi^-$ process. The 
cross sections for the elementary reaction were measured in the 1960s and early 
1970s using hydrogen bubble chambers - the total cross section data from these 
measurements are tabulated in Ref.~\cite{fla83}. It is essential to investigate 
first the $(K^-,K^+)$ reaction on a proton target leading to the production of a 
free $\Xi^-$ hyperon. The input information extracted from this study will then 
be used in the description of the formation of $\Xi^-$-hypernuclei using this 
reaction.

In this paper, we investigate the production of free cascade hyperon and cascade 
hypernuclei via the $(K^-,K^+)$ reaction on proton [Fig.~1(a)] and nuclear targets 
[Fig.~1(b)], respectively  within an effective Lagrangian model~\cite{shy99,shy09}. 
It retains the full field theoretic structure of the interaction vertices and 
treats baryons as Dirac particles. The initial state interaction of the incoming 
$K^-$ with a free or bound target proton leads to excitation of intermediate 
$\Lambda$ and $\Sigma$ resonant states, which propagate and subsequently decay into 
$\Xi^- K^+$. In case of the reaction on nuclei, $\Xi^-$  gets captured into one 
of the nuclear orbits, while the $K^+$ meson goes out.  We have considered the 
intermediate resonant states, $\Lambda$, $\Sigma$ and eight of their resonances 
with masses up to 2.0 GeV [$\Lambda(1405)$, $\Lambda(1520)$, $\Lambda(1670)$, 
$\Lambda(1810)$, $\Lambda(1890)$, $\Sigma(1385)$, $\Sigma(1670)$, $\Sigma(1750)$], 
which are represented by $\Lambda^*$ and $\Sigma^*$ in Figs.~1(a) and 1(b).

\begin{figure}[htb]
\begin{center}
\includegraphics[width=.25\textwidth]{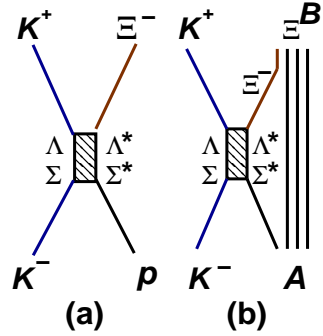}
\end{center}
\vspace{-0.5cm}
\caption{Graphical representation of the model used to describe $p(K^-, K^+)\Xi^-$
(a) and $A(K^-, K^+){_{\Xi^-}}\!\! B$ (b) reactions. $A$ represents the target nucleus 
and ${_{\Xi^-}}\!\! B$ the final hypernucleus.}
\label{fig1}
\end{figure}

\section{Formalism}

The differential cross section for the $(K^-, K^+)$ reaction is given by
\begin{equation}
        d\sigma = \frac{1}{(2\pi)^2}
        \, \frac{d^3 p_{K^+}}{2E_{K^+}}
        \, \frac{d^3 p_{B}}{2E_{B}}
        \, \frac{m_{A} m_{B}}{|{\bf p}_{K^-}|\sqrt{s}}
        \, |\sum_{R_i} \mathcal{M}_{R_i}|^2
        \delta^{(4)}\left( p_{K^-} + p_{A} - (p_{K^+} + p_{B}) \right)
        \;.
\end{equation}
The summation over initial ($m_i$) and final ($m_f$) spin states is implied. 
$\sum_{R_i}$ indicates the summation over all the resonance intermediate states. 
In Eq.~1, A and B represent the masses of the target and the residual nuclei, 
respectively. In case of the elementary reaction, they are replaced by the proton 
mass and the cascade mass, respectively.

In order to calculate the amplitude $\mathcal{M}_{R_i}$ for the elementary $\Xi$
hyperon production reaction, one requires the effective Lagrangians at the 
meson-baryon-resonance vertices (which involve coupling constants and the 
form factors), and the propagators for various resonances. They were taken to 
be the same as those given Refs.~\cite{shy11,shy12}. In addition, for the 
hypernuclear production reactions one needs spinors for the nucleon hole and 
$\Xi^-$ particle bound states. We have used a phenomenological model to get 
them as discussed in the next section.

\section{Results and discussions} 

In Figs.~2(a), we show comparisons of our calculations with the data for the total 
cross section of the $p(K^-, K^+)\Xi^-$ reaction as a function of $K^-$ beam
momenta ($p_{K^-}$). It is clear that our model reproduces the data reasonably  
well within statistical errors.  The measured total cross section peaks in the 
region of 1.4-1.5 GeV/c which is well described by our model. In Fig.~2(b), we 
show the ($p_{K^-}$) dependence of the $0^\circ$ differential cross section 
($d\sigma/d\Omega)_{0^\circ}$ for the $p(K^-,K^+)\Xi^-$ reaction. This quantity is 
of interest as it enters explicitly into the expression for the cross sections of 
the $A(K^-, K^+){_{\Xi^-}}\!\! B$ reactions within the  impulse approximation models 
(see, e.g. Ref.~\cite{shy12}).  It may be noted that we have not normalized the 
cross sections in Fig.~2(b) in the way it was done in the similar results shown in 
Fig.~5 of Ref.~\cite{shy12}. We see that [($d\sigma/d\Omega)_{0^\circ}$] peaks in 
the same region of $p_{K^-}$ as the total cross section shown in Fig.~2a.  The 
arrows in Figs.~2(a) and 2(b) show the position of the threshold beam momentum 
for this reaction. 
\begin{figure}[t]
\begin{center}
\includegraphics[width=.25\textwidth]{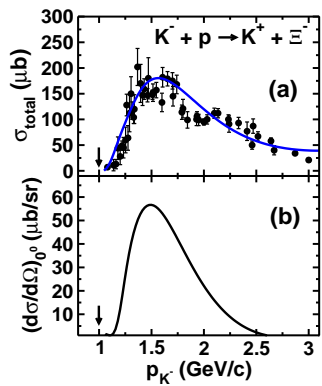}
\end{center}
\vspace{-0.5cm}
\caption{(a) Total cross section for the $p(K^-, K^+)\Xi^-$ reaction as a 
function of incident $K^-$ momentum; the experimental data is from 
~\protect~\cite{fla83}. (b) The zero degree differential cross section of 
the $p(K^-, K^+)\Xi^-$ reaction as a function of beam momentum. }
\label{fig2}
\end{figure}

Next, we discuss the hypernuclear production reactions
 $^{12}$C$(K^-,K^+)^{12}{\!\!\!_{\Xi^-}}$Be and 
$^{28}$Si$(K^-,K^+)^{28}{\!\!\!_{\Xi^-}}$Mg. The thresholds for these reactions 
are about 0.761 GeV/c and 0.750 GeV/c, respectively. We have employed pure 
single-particle-single-hole $(p^{-1}\Xi)$ configurations to describe
the nuclear structure part. For the proton hole and $\Xi^-$ states, spinors 
were generated in a phenomenological model, where they are obtained by solving 
the Dirac equation with scalar and vector fields having Woods-Saxon radial forms. 
With a set of radius and diffuseness parameters, the depths of these fields 
($V_s$ and $V_v$, respectively) are searched to reproduce the binding energy (BE) 
of a given state. Since the experimental values of the BEs for the $\Xi^-$ bound 
states are as yet unknown, we have adopted in our search procedure the predictions 
for BEs  made in the latest version of the quark-meson coupling (QMC) model
\cite{gui08}. This model predicts only one bound state for the 
$^{12}{\!\!\!_{\Xi^-}}$Be hypernucleus with a  binding energy of 3.038 MeV and 
quantum numbers ($1s_{1/2}$). On the other hand, for $^{28}{\!\!\!_{\Xi^-}}$Mg 
it predicts 3 distinct bound $\Xi^-$ states, $1s_{1/2}$, $1p_{3/2}$, $1p_{1/2}$, 
with corresponding binding energies of 8.982 MeV, 4.079 MeV, and 4.414 MeV, 
respectively~\cite{kaz12}. The values of radius and diffuseness parameters were 
taken to be 0.983 fm and 0.606 fm, respectively for both scalar and vector fields. 
The derived values of ($V_s$, $V_v$) were (all in MeV) (-112.11, 90.81), 
(-133.53, 108.16), (-188.98, 153.08) and (-205.70, 166.65) for above four states, 
respectively. It may be noted that these values are obtained without including
the Coulomb interaction between $\Xi^-$ and $^{11}$B. $V_s$+ $V_v$ should be 
comparable to the depth of the corresponding conventional Woods-Saxon potential. 

In case of the $^{12}$C target, the $\Xi^-$ hyperon in a 1$s_{1/2}$ state can 
populate 1$^-$ and 2$^-$ states of the hypernucleus corresponding to the 
particle-hole configuration $[(1p_{3/2})^{-1}_p,(1s_{1/2})_{\Xi^-}$]. The states 
populated for the $^{28}{\!\!\!_{\Xi^-}}$Mg hypernucleus are [2$^+$, 3$^+$], 
[1$^-$, 2$^-$, 3$^-$, 4$^-$], and [$2^-$, $3^-$] corresponding to the 
configurations $[(1d_{5/2})^{-1}_p, (1s_{1/2})_{\Xi^-}$], $[(1d_{5/2})^{-1}_p, 
(1p_{3/2})_{\Xi^-}$], and $[(1d_{5/2})^{-1}_p, (1p_{1/2})_{\Xi^-}$], respectively. 
In Fig.~3, we have shown results for populating the hypernuclear state with 
maximum spin of natural parity for each configuration.  We have used a plane 
wave approximation to describe the relative motion of kaons in the incoming and 
outgoing channels. However, distortion effects are partially accounted for by 
introducing reduction factors to the cross sections as described in Refs.
\cite{dov83,ike94}. It should be noted that the cross sections obtained with the 
spinors calculated within the QMC model are very close to those shown in Fig.~3
\cite{shy12}. It is also of interest to note that the cross sections shown in both
Figs.~2 and 3 are dominated by the contributions of the $\Lambda(1520)$ ($D_{03}$)  
resonance intermediate state.

It is clear from Fig.~3 that for both the hypernuclear production reactions, the 
cross sections peak at $p_{K^-}$ around 1.0 GeV/c which is $\approx$ 0.3 GeV/c above 
the production thresholds of the two reactions. It is not too different from the 
case of the elementary $\Xi^-$ production cross sections where the peaks of the 
cross sections occur at about 0.35 -0.40 GeV/c above the corresponding production
threshold (see Fig~2). The magnitudes of the cross sections for the 
$^{12}{\!\!\!_{\Xi^-}}$Be production are in excess of 1 $\mu b$ near the 
peak position.  It is important to note that at the beam momentum of 1.6 
GeV/c, the magnitude of our cross section for this case is similar to that 
obtained in Ref.~\cite{ike94} within an impulse approximation model. Moreover, 
our cross sections at 1.8 GeV/c also are very close to those of Ref.~\cite{dov83} 
for both the targets. A more rigorous consideration of the distortion effects could 
alter the pattern of the beam momentum dependence, {\it e.g.} it is likely to be 
relatively stronger at lower values of $p_{K^-}$ as compared to higher values. 
Furthermore, Coulomb interactions between $\Xi^-$ and the core nucleus may also 
have some effect for heavier targets. These effects will be investigated in a future 
publication. 
\begin{figure}[t]
\begin{center}
\includegraphics[width=.25\textwidth]{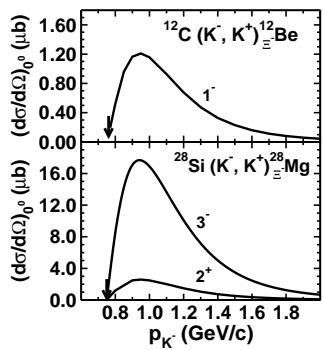}
\end{center}
\vspace{-0.4cm}
\caption{Differential cross section at $0^\circ$ as a function of
$K^-$ beam momentum for the $^{12}$C$(K^-,K^+)^{12}{\!\!\!_{\Xi^-}}$Be
and $^{28}$Si$(K^-,K^+)^{28}{\!\!\!_{\Xi^-}}$Mg reactions. The spin-parity
of the final hypernuclear states are indicated on each curve. The solid 
lines represent the cross sections obtained with the phenomenological $\Xi^-$ 
bound state spinors. Arrows show the threshold for the corresponding reaction.}
\label{fig3}
\end{figure}

\section{Summary and conclusions}

In summary, we studied the production of $\Xi^-$ hyperon and the corresponding 
hypernuclei in the $(K^-,K^+)$ reaction on proton and nuclear targets, respectively 
within an effective Lagrangian model where the reaction proceeds via excitation, 
propagation and decay of $\Lambda$ and $\Sigma$ hyperon resonance states  in the 
initial collision of $K^-$ meson with the initial free or bound proton. In the 
calculations of $\Xi^-$ hypernuclear production the bound state spinors were 
obtained within a phenomenological model. The data on the total cross section of 
the elementary production reaction are well described by our model. The zero degree 
differential cross sections for the $\Xi^-$ hypernuclear production peak around the 
beam momentum of 1.0 GeV/c with peak cross section of more than 1 $\mu b$. They 
closely follow the trends of  the elementary $\Xi^-$ production cross sections. Our 
predictions will be useful for the future JPARC experiments.  

\section{Acknowledgments}
The author wishes to thank O. Scholten, K. Tsushima and A.W. Thomas for several 
useful discussions.

\end{document}